\documentclass[onecolumn,showpacs,preprintnumbers,amsmath,amssymb]{revtex4}

\usepackage{graphicx}
\usepackage{dcolumn}
\usepackage{bm}
\begin{document}
\newcommand{\hs}{\hspace*{0.5cm}}
\newcommand{\vs}{\vspace*{0.5cm}}
\newcommand{\be}{\begin{equation}}
\newcommand{\ee}{\end{equation}}
\newcommand{\bea}{\begin{eqnarray}}
\newcommand{\eea}{\end{eqnarray}}
\newcommand{\ben}{\begin{enumerate}}
\newcommand{\een}{\end{enumerate}}
\newcommand{\bde}{\begin{widetext}}
\newcommand{\ede}{\end{widetext}}
\newcommand{\nn}{\nonumber}
\newcommand{\crn}{\nonumber \\}
\newcommand{\non}{\nonumber}
\newcommand{\noi}{\noindent}
\newcommand{\al}{\alpha}
\newcommand{\la}{\lambda}
\newcommand{\bet}{\beta}
\newcommand{\ga}{\gamma}
\newcommand{\va}{\varphi}
\newcommand{\om}{\omega}
\newcommand{\pa}{\partial}
\newcommand{\fr}{\frac}
\newcommand{\bc}{\begin{center}}
\newcommand{\ec}{\end{center}}
\newcommand{\Ga}{\Gamma}
\newcommand{\de}{\delta}
\newcommand{\De}{\Delta}
\newcommand{\ep}{\epsilon}
\newcommand{\varep}{\varepsilon}
\newcommand{\ka}{\kappa}
\newcommand{\La}{\Lambda}
\newcommand{\si}{\sigma}
\newcommand{\Si}{\Sigma}
\newcommand{\ta}{\tau}
\newcommand{\up}{\upsilon}
\newcommand{\Up}{\Upsilon}
\newcommand{\ze}{\zeta}
\newcommand{\ps}{\psi}
\newcommand{\Ps}{\Psi}
\newcommand{\ph}{\phi}
\newcommand{\vph}{\varphi}
\newcommand{\Ph}{\Phi}
\newcommand{\Om}{\Omega}
\def\lappeq{\mathrel{\rlap{\raise.5ex\hbox{$<$}}
{\lower.5ex\hbox{$\sim$}}}}
\vspace*{0.5cm}
\title{ Study of the singly charged Higgs in the economical 3-3-1  model at $e^+e^-$colliders}

\author{Dang Van Soa}
\author{Dao Le Thuy}
\email{dvsoa@assoc.iop.vast.ac.vn} \affiliation{Department of
Physics, Hanoi University of Education, Hanoi, Vietnam}

\date{\today}
\begin{abstract}
The  $\mathrm {SU}(3)_C\otimes \mathrm {SU}(3)_L \otimes {\mathrm
U}(1)_X$ gauge model with two Higgs triplets ( the economical
3-3-1 model) are presented. This model predicts a new
singly-charged Higgs bosons $H_2^\pm$, namely a scalar bilepton (
a particle of lepton number 2 ) with  mass proportional to the
bilepton mass $M_Y$ through the coefficient $\la_4$. Pair
production of $H_2^\pm$ at high energy $e^+ e^-$ colliders
 with polarization of $e^+$, $e^-$ beams was studied in detail.
 Numerical evaluation shows that
if Higgs mass is not too heavy and at high degree of polarization
then the reaction can give observable cross section in future
colliders. The reaction $e^+ e^- \rightarrow H^{-}_2W^+ $ is also
examined. Based on the result, it  shows that production cross
sections of $H^{-}_2W^+$ are very small, much below
 pair production of $H^{\pm}_2$, so that the direct
production of them is in general not expected to lead to
easily observable signals in $e^+e^-$ annihilation.\\

\end{abstract}


\maketitle

\section{Introduction}
\label{Intro}

\hspace*{0.5cm}Recent neutrino  experimental results~\cite{superk}
establish the fact that neutrinos have masses and the standard
model (SM) must be extended. Among the beyond-SM extensions, the
models based on the $\mbox{SU}(3)_C\otimes \mbox{SU}(3)_L \otimes
\mbox{U}(1)_X$ (3-3-1) gauge group have some  intriguing features:
Firstly, they can give partial explanation of  the generation
number problem. Secondly,  the third quark generation has to be
different from the first two, so this leads to the possible
explanation of why top quark is uncharacteristically heavy.

\hspace*{0.5cm}There are two main versions of the 3-3-1 models. In
one of them \cite{ppf} the three known left-handed lepton
components for each generation are associated to three
$\mathrm{SU}(3)_L$ triplets as $(\nu_l,l,l^c)_L$, where $l^c_L$ is
related to the right-handed isospin singlet of the charged lepton
$l$ in the SM. The scalar sector of this model is quite
complicated (three  triplets and one sextet). In the variant model
\cite{flt} three $\mathrm{SU}(3)_L$ lepton triplets are of the
form $(\nu_l, l, \nu_l^c)_L$, where $\nu_l^c$ is related to the
right-handed component of the neutrino field $\nu_l$ (a model with
right-handed neutrinos). The scalar sector of this model requires
three Higgs triplets, therefore, hereafter we call this version
the 3-3-1 model with three Higgs triplets (331RH3HT). It is
interesting to note that, in the 331RH3HT,  two Higgs triplets
have the same $\mathrm{U}(1)_X$ charge with two neutral components
at their top and bottom. Allowing these neutral components vacuum
expectation values (VEVs), we can reduce number of Higgs triplets
to be two. Based on the results, the model with two higgs
triplets, namely {\it the economical 3-3-1 model} was proposed
recently~\cite{haihiggs}. This model contains very important
advantage, namely: There is no new parameter, but it contains very
simple Higgs sector, hence the significant number of free
parameters is reduced. In the Higgs sector of this model there are
four physical Higgs bosons. Two of them are neutral physical
fields($H^0$ and $H^0_1$), others are singly charged bosons
$H_2^\pm$.\\
\hspace*{0.5cm} The Higgs boson plays an important role in the SM,
it is responsible for generating the masses of all the elementary
particles ( leptons, quarks, and gauge bosons ). The experimental
 detection of them will be great triumph of the electroweak
 interactions and will mark new stage in high energy physics. However,
the mass of the Higgs boson is a free parameter~\cite{gun}.
 The trilinear Higgs self-coupling can
be measure directly in pair production of Higgs particle at hadron
and high energy $e^+e^-$ linear colliders~\cite{ily,daw, bae,eac},
or at photon colliders~\cite{jik}. Interactions among the standard
model gauge bosons and Higgs bosons in the economical 3-3-1 model
were studied in detail in our early work~\cite{haihiggs1}. The
trilinear gauge boson couplings in the 3-3-1 models were presented
in~\cite{longs} and production of bileptons in high energy
collisions was studied recently~\cite{soa}. Discovery potential
for doubly charged Higgs bosons and the possibility to detect the
neutral Higgs boson in the minimal version  at $e^+ e^-$ colliders
was considered ~\cite{step,cien}. Production of the Higgs bosons
in the 331RH3HT at the CERN LHC has been done in Ref.~\cite{ninhlong}.\\
\hspace*{0.5cm}The polarization of electron and positron beams
gives a very effective means to control the effect of the SM
processes on the experimental analyses. Beam polarization is also
an indispensable tool for the identification and study of new
particles and their interactions. In this paper we turn our
attention to the future perspective experiment, namely the
production of the singly charged Higgs in the economical 3-3-1
model at  high energy collisions
 with polarization of $e^+$, $e^-$ beams.
The rest of this  paper is organized as follows: In Section
\ref{model}, we give a brief review of the economical 3-3-1 model.
Section \ref{charged} represents a detail calculation for
 cross section of pair production $H_2^\pm$ with polarization of
$e^+$, $e^-$ beams. Pair production of $H^-_2W^+$ via $ZZ'$ fusion
is given in Sec. \ref{charged1}. We summarize our result and make
conclusions in the last section - Sec. \ref{conclus}.
\section{A review of the economical 3-3-1  model }
\label{model}
 The particle content in this model, which is anomaly
free, is given as follows~\cite{haihiggs}
 \be \psi_{aL} = \left(
               \nu_{aL}, l_{aL}, N_{aL}
\right)^T \sim (1, 3, -1/3),\hs l_{aR}\sim (1, 1, -1),
  \label{l2}
\vspace*{0.5cm} \ee where $a = 1, 2, 3$ is a  family index. Here
the right-handed neutrino is denoted by $ N_L \equiv (\nu_R)^c$.
  \bea
 Q_{1L}&=&\left( u_1,  d_1,  U \right)^T_L\sim
 \left(3,\fr 1 3\right),\hs Q_{\al L}=\left(
  d_\al,  -u_\al,  D_\al
\right)^T_L\sim (3^*,0),\hs \al=2,3,\crn u_{a R}&\sim&\left(1,\fr
2 3\right), \hs d_{a R} \sim \left(1,-\fr 1 3\right), \hs
U_{R}\sim \left(1,\fr 2 3\right),\hs D_{\al R} \sim \left(1,-\fr 1
3\right).\eea  Electric charges of the exotic quarks $U$ and
$D_\al$  are the same as of the usual quarks, i.e. $q_{U}=\fr 2 3$
and $q_{D_\al}=-\fr 1 3$. The $\mathrm{SU}(3)_L\otimes
\mathrm{U}(1)_X$ gauge group is broken spontaneously via two
steps. In the first step, it is embedded in that of the SM via a
Higgs scalar triplet \be \chi=\left(  \chi^0_1,  \chi^-_2,
\chi^0_3 \right)^T \sim \left(3,-\fr 1 3\right),\ee acquired with
VEV given by \be \langle\chi\rangle=\fr{1}{\sqrt{2}}\left(  u,  0,
\om \right)^T.\label{vevc}\ee In the last step, to embed the gauge
group of the SM in $\mathrm{U}(1)_Q$, another Higgs scalar triplet
\be \phi=\left(   \phi^+_1,  \phi^0_2,  \phi^+_3 \right)^T\sim
\left(3,\fr 2 3\right),\ee is needed with the VEV as follows \be
\langle\phi\rangle =\fr{1}{\sqrt{2}}\left(
  0,  v,  0 \right)^T.\label{vevp}\ee
\label{charged} The Yukawa interactions which induce masses for
the fermions can be written in the most general form as \be
\mathcal{L}_Y=(\mathcal{L}^\chi_Y+\mathcal{L}^\phi_Y)+\mathcal{L}^{\mathrm{mix}}_Y,\ee
where \bea ({\cal L}^{\chi}_Y+{\cal
L}^\phi_Y)&=&h'_{11}\overline{Q}_{1L}\chi
U_{R}+h'_{\al\beta}\overline{Q}_{\al L}\chi^* D_{\beta R}\crn
&&+h^e_{ij}\overline{\psi}_{iL}\phi
e_{jR}+h^\ep_{ij}\ep_{pmn}(\overline{\psi}^c_{iL})_p(\psi_{jL})_m(\phi)_n
\crn &&+h^d_{1i}\overline{Q}_{1 L}\phi d_{i R}+h^d_{\al
i}\overline{Q}_{\al L}\phi^* u_{iR}+h.c.,\label{y1}\\ {\cal
L}^{\mathrm{mix}}_Y&=&h^u_{1i}\overline{Q}_{1L}\chi
u_{iR}+h^u_{\al i}\overline{Q}_{\al L}\chi^* d_{i R}\crn
&&+h''_{1\al}\overline{Q}_{1L}\phi D_{\al R}+h''_{\al
1}\overline{Q}_{\al L}\phi^* U_{R}+h.c.\label{y2}\eea
 The VEV $\om$ gives mass for the exotic quarks $U$, $D_\al$ and the new
gauge bosons $Z^{\prime},\ X,\ Y$, while the VEVs $u$ and $v$ give
mass for the quarks $u_a,\ d_a$, the leptons $l_a$ and all the
ordinary gauge bosons $Z,\ W$. To keep a consistency with the
effective theory, the VEVs in this model have to satisfy the
constraint \be u^2 \ll v^2 \ll \om^2. \ee In this model, the most
general Higgs potential has very simple form \bea V(\chi,\phi) &=&
\mu_1^2 \chi^\dag \chi + \mu_2^2 \phi^\dag \phi + \la_1 (
\chi^\dag \chi)^2 + \la_2 ( \phi^\dag \phi)^2\crn & & + \la_3 (
\chi^\dag \chi)( \phi^\dag \phi) + \la_4 ( \chi^\dag \phi)(
\phi^\dag \chi). \label{poten} \eea Note that there is no
trilinear scalar coupling and this makes the Higgs potential much
simpler than those in the 331RN3HT \cite{flt} and closer to that
of the SM. The analysis in Ref.\cite{study} shows that after
symmetry breaking, there are eight Goldstone bosons and four
physical scalar fields. One of two physical neutral scalars is the
SM  Higgs boson.
 In the pseudoscalar
sector, there are two neutral physical fields-the SM $H^0$ and the
new $H^0_1$ with masses.
\bde \bea M^2_{H^0}\simeq\fr{4\la_1\la_2-\la^2_3}{2\la_1}v^2,\label{potenn10a}\\
M^2_{H^0_1}\simeq 2\la_1 \om^2.\label{potenn10}\eea \ede From
Eq.(\ref{potenn10}), it follows that mass of the new Higgs boson
$M_{H^0_1}$ is related to mass of the bilepton gauge $X^0$ (or
$Y^\pm$ via the law of Pythagoras) through \bea M^2_{H^0_1}& =&
\fr{8\la_1}{g^2} M_X^2 \left[1 +
\mathcal{O}\left(\fr{M_W^2}{M_X^2}\right)\right]\crn &=& \fr{2
\la_1 s^2_W}{\pi \al} M_X^2  \left[1 +
\mathcal{O}\left(\fr{M_W^2}{M_X^2}\right)\right] \approx 18.8
\la_1 M_X^2. \label{potenn13}\eea Here, we have used $\al
=\textbf{}
\fr{1}{128}$ and $s^2_W = 0.231$.\\
In the charged Higgs sector,there are two charged Higgs bosons
$H_2^\pm$ with mass \bea M^2_{H_2^\pm}&
=&\fr{\la_4}{2}(u^2+v^2+\om^2) = 2 \la_4 \fr{M_Y^2}{g^2}\crn & = &
\fr{s_W^2 \la_4}{2 \pi \al} M_Y^2 \simeq 4.7 \la_4
M_Y^2.\label{potenn14}\eea

From Eq. (\ref{potenn13}) we see that the  masses of the new
neutral Higgs boson $H^0_1$ and the neutral non-Hermitian bilepton
$X^0$ are dependent on a coefficient of Higgs self-coupling
($\la_1$).
 Eq.(\ref{potenn14}) gives us a connection between  mass of $H^\pm_2$ and
 bilepton mass $Y$ through the coefficient of Higgs self-coupling
$\la_4$. Note that in the considered model.
 To keep the smallness of these couplings,
 the mass $M_{H^\pm_2}$ can be taken
 in the electroweak scale with $\la_4\sim 0.01$~\cite{keu}.
From (\ref{potenn14}), taking the lower limit for $M_Y$ to be 1
TeV, the mass of $H^\pm_2$ is in range of 200 GeV.
 Interactions among the standard model gauge bosons and Higgs
bosons were studied in detail~\cite{haihiggs1}. From these
couplings, all scalar fields including the neutral scalar $H^0$
and the Goldstone bosons were identified and their couplings with
the usual gauge bosons such as the photon, the charged bosons
$W^\pm$ and the neutral $Z$, and also $Z'$ without any additional
condition were recovered. Note that The CP-odd part of Goldstone
associated with the neutral non-Hermitian bilepton gauge boson
$G_{X^0}$ is decouple, while its CP-even counterpart has the
mixing by the same way in the gauge boson sector.\\
 \section{ Pair production of $H^\pm_2$ in $e^+ e^-$ colliders }
\label{charged}

\hspace*{0.5cm} High energy $e^+$$e^-$ colliders have been
essential instruments to search for the fundamental constituents
of matter and their interactions. The possibility to detect the
neutral Higgs boson in the minimal version  at $e^+ e^-$ colliders
was considered in~\cite{cien} and production of the SM-like
neutral Higgs boson at the CERN LHC was studied in
Ref.\cite{ninhlong}. This section is devoted to pair production of
$H^\pm_2$ at $e^+ e^-$ colliders with the center-of-mass energy is
chosen in range between  $500 GeV$ (ILC) and $1000 GeV$. The
trilinear couplings of $H^\pm_2$ with the neutral gauge bosons in
the SM are given in Table I \cite{haihiggs1}
\begin{table}[h]
\caption{The trilinear couplings of the pair $H_2^\pm$ with the
neutral gauge bosons in the SM.} \bc
\begin{ruledtabular}
\begin{tabular}{c|ccc}
Vertex  & $A^\mu H^-_2\overleftrightarrow{\pa_\mu}H^+_2$ & $Z^\mu
H^-_2\overleftrightarrow{\pa_\mu}H^+_2$ \\ \hline \\
Coupling & $i e$ & $-ig s_{W}t_{W}$ &
\end{tabular}
\end{ruledtabular}
 \label{tab6}
\ec
\end{table}

From Table I. we can see that pair production of $H^\pm_2$ in $e^+
e^-$ colliders exists through the neutral gauge bosons in the SM.
Now we consider the process in which the initial state contains
the electron and the positron and in the final state there are the
pair $H^\pm_2$  \be e^-(p_1) + e^+(p_2)
\rightarrow H^+ _2(k_1) + H^-_ 2(k_2), \ee \\
The straightforward calculation yields the following differential
cross section (DCS) in the center-of-mass frame with the polarized
$e^+$,$e^-$ beams is
\begin{eqnarray}
\frac{d\sigma_{(P_+, P_-)}}{d \cos \theta}
&=&\left(\frac{K^3_{H^+_2H^-_2}\pi\alpha^2}{4
s^2}\right)\Big[\frac{1}{s^2}(1-P_+ P_-) - \frac{g_z
[v_e(1-P_+P_-)-a_e(P_- - P_+)]}{2 C_W S_W s ( s-m^2_Z)}\crn && +
\frac{g^2_z [(v^2_e+ a^2_e)(1-P_- P_+)-2v_ev_a(P_- - P_+)]}{16
C^2_W S^2_W(
s-m^2_Z)^2}\Big]\left(1-cos^2\theta\right),\label{eq.21}
\end{eqnarray}
where $\theta$ is angle between $\overrightarrow{p_1}$ and
$\overrightarrow{k_1}$, $g=\frac{e}{s_W}$; $g_Z$=$\frac{1}{2S_W
C_W(\omega^2+v^2)}(2\omega^2 S^2_{W}-v^2(4 C^2_{W} -1))$; $v_e=
-1+ 4 s^2_W$;\\ $a_e= -1$,  and
\begin{eqnarray}
K_{H^+_2H^-_2}=\left[(s-m^2_{H^+_2} - m^2_{H^-_2})^2 - 4
m^2_{H^+_2} m^2_{H^-_2}\right]^{1/2}.
\end{eqnarray}
 $P_+$ and $P_-$ are
polarized coefficients of $e^+$ and $e^-$ beams, respectively.
The total cross section is given by\\
\begin{eqnarray}
\sigma_{(P_+, P_-)}&=&\left(\frac{K^3_{H^+_2H^-_2}\pi\alpha^2}{3
s^2}\right)\Big[\frac{1}{s^2}(1-P_+ P_-) - \frac{g_z
[v_e(1-P_+P_-)-a_e(P_- - P_+)]}{2 C_W  S_Ws ( s-m^2_Z)}\crn && +
\frac{g^2_z [(v^2_e+ a^2_e)(1-P_- P_+)-2v_ev_a(P_- - P_+)]}{16
C^2_W S^2_W( s-m^2_Z)^2}\Big]\label{eq.22},
\end{eqnarray}
 here we take $m_Z=91.1882$ $GeV$, $v = 246$  $GeV$
$\omega=1TeV$ and $m_H=200$ $GeV$~\cite{haihiggs}.

\begin{table}[h]
\caption{Number of events with the integrated luminosity of $500
fb^{-1}$ and $m_{H} = 200 GeV$}
\begin{ruledtabular}
\begin{tabular}{lccc}
 $\sqrt{s}$ (GeV) & $500$ & $750$ & $1000$ \\
 \hline \\ $\sigma_{(P_+, P_-)}$ ($pbarn$) & $ 8.7\times 10^{-2}$ & $7.4\times 10^{-2}$ & $6.1\times 10^{-2}$\\
\hline \\ Number of events & $4350$ & $3700$ & $3050$
\end{tabular}
 \end{ruledtabular}
\end{table}

In Fig.1 we have plotted the cross section $\sigma_{(P_+, P_-)}$
as a function of the $P_+$ and $P_-$. The collision energy is
taken to be  $\sqrt{s} = 1$$TeV$, and the Higgs  mass
$m_{H^{\pm}_2} = 200$$GeV$. The figure shows that  the cross
section gets a maximum value as $\sigma_{(P_+, P_-)}$ = 6$\times
10^{-2}$ $ pbarn$ for $P_- = -1$ and $P_+ = 1$, while $\sigma_0$ =
2.1$\times 10^{-2}$ $pbarn$ for $P_- = P_+ = 0$. The DCS as
function of $cos\theta$ is shown in Fig.2, the curve 1 for
$P_-=P_+=0$ and the curve 2 for $P_-=-1$, $P_+=1$. From Fig.2 we
can see that the DCS gets a maximum  value at $cos\theta=0$, which
is similar to the process $e^+ e^- \rightarrow  Z h$ in the SM  (
while the DCS of
 the reactions $e^+ e^- \rightarrow Z Z, Z A $ gets the minimal value at
 $cos\theta=0$ )\cite{eac}.
The cross section with highly polarized $e^+$, $e^-$ beams is much
larger than those with non-polarized $e^+$, $e^-$ beams.
Following, the dependence of the total cross-section on Higgs mass
 for the fixed collision energies, typically $\sqrt{s}$ = $0.5TeV$,
 $\sqrt{s}$ = $0.75TeV$ and $\sqrt{s}$ = $1TeV$, is also shown in Fig. 3 for
 $P_-=-1$, $P_+=1$ and in Fig. 4 for $P_-=P_+=0$. The Higgs mass is chosen as
$200$ $GeV$ $\leq m_H \leq 500$ $GeV$. As we can see from the
figures, at high energies, the $m_H$ increases then the
cross-section decreases, which is equal to zero at $m_H \simeq
500$ $GeV$ for $\sqrt{s} = 1$ $TeV$, at $m_H \simeq 440$ $GeV$ for
$\sqrt{s}$ = $\sqrt{0.75}$$TeV$, and at $m_H \simeq 350GeV$ for
$\sqrt{s}$ = $\sqrt{0.5}$$TeV$. It is worth noting that there is
no difference between two lines at $m_H \simeq 230$ $GeV$, $m_H
\simeq 250$ $GeV$ and $m_H \simeq 300$ $GeV$. Taking the
integrated luminosity of $500 fb^{-1}$~\cite{bae}, Higgs mass is
chosen at the relatively low value of $200 GeV$,  the number of
events with different values of $\sqrt{s}$ is given in Table II.
From the results , it shows that with high integrated luminosity
and at high degree of polarization of electron and positron beams
then the production cross section may give observable values at
moderately high energies.
The number of events can be several thousand as expected.\\

\section{
 Production of $H^-_2W^+$ via $ZZ'$ fusion at $e^+e^-$ colliders}
\label{charged1}

The similarity as section III, in this section we examine the pair
production of $H^-_2W^+$ at $e^+ e^-$ colliders via $Z$ $ Z'$
exchange. The trilinear couplings of the pair $H^-_2W^+$ with the
neutral gauge bosons in the economical 3-3-1 model are given
by\cite{haihiggs1}
\begin{table}[h]
\caption{The trilinear couplings of the pair $H^-_2W^+$ with the
neutral gauge bosons in the economical 3-3-1 model} \bc
\begin{ruledtabular}
\begin{tabular}{c|ccc}
Vertex  & $Z^\mu H^-_2\pa_\mu W^+$ & $Z'^\mu
H^-_2\pa_\mu W^+$ \\ \hline \\
Coupling & $\frac{g^2 v \omega}{2\sqrt{\omega^2+c^2_\theta
v^2}}[s_\theta c_\theta(u_{12}+\sqrt{3} u_{22})+c_{2\theta}
u_{42}]$ & $\frac{g^2 v \omega}{2\sqrt{\omega^2+c^2_\theta
v^2}}[s_\theta c_\theta(u_{13}+\sqrt{3} u_{23})+c_{2\theta}
u_{43}]$ &
\end{tabular}
\end{ruledtabular}
 \label{tab6}
\ec
\end{table}

Here $U_{\al\beta}$ ($\al, \beta = 1,2,3,4$ ) is a mixing matrix
of neutral gauge bosons given in appendix A. The angle $\theta$ is
a mixing angle between charged gauge bosons $W$ and $Y$, which is
defined by ~\cite{haihiggs}\be \tan\theta=\fr{u}{\om}.\ee The
decay of charged gauge boson $W$ into leptons and quarks was
analyzed in detail in~\cite{haihiggs}, which  gives us a upper
limit: $\sin \theta \leq 0.08$. From Table II. we can see that
pair production of $H^- _2 W^+$ in $e^+ e^-$ colliders exists
through the neutral gauge bosons $Z$ and $Z$' in
the $s$- channel\\
\be e^-(p_1) + e^+(p_2)
\rightarrow H^- _2(k_1) + W^+(k_2), \ee \\
In this case the DCS is given by

\begin{eqnarray}
\frac{d\sigma_{(P_+, P_-)}}{d \cos \theta}
&=&\left(\frac{K_{H^-_2W^+}\pi^2\alpha^3}{2
s^2}\right)\Bigg\{\frac{e^2_Z[(v^2_e+a^2_e)(1-P_- P_+)-2v_e
a_e(P_- - P_+)]}{16C^2_W S^2_W(s-m_Z)^2}\crn && +
\frac{e'^2_Z[(v'^2_e+a'^2_e)(1-P_- P_+)- 2v'_e a'_e(P_- -
P_+)]}{16C^2_W S^2_W(s-m_{Z'})^2}\crn && + \frac{e_Z e'_Z[(v_e
v'_e+a_ea'_e)(1-P_- P_+)-(v_e a'_e+v'_e a_e)(P_- - P_+)]}{8C^2_W
S^2_W(s-m^2_Z)(s-m^2_{Z'})}\Bigg \}\left\{-2s+\frac{K^2_{H^-_2
W^+}}{2m^2_W}(1-cos^2\theta)\right\},\label{eq.23}
\end{eqnarray}
and  the total cross-section is
\begin{eqnarray}
\sigma_{(P_+, P_-)}&=&\left(\frac{K_{H^-_2W^+}\pi^2\alpha^3}{2
s^2}\right)\Bigg\{\frac{e^2_Z[(v^2_e+a^2_e)(1-P_- P_+)-2v_e
a_e(P_- - P_+)]}{16C^2_W S^2_W(s-m_Z)^2}\crn && +
\frac{e'^2_Z[(v'^2_e+a'^2_e)(1-P_- P_+)- 2v'_e a'_e(P_- -
P_+)]}{16C^2_W S^2_W(s-m_{Z'})^2}\crn && + \frac{e_Z e'_Z[(v_e
v'_e+a_ea'_e)(1-P_- P_+)-(v_e a'_e+v'_e a_e)(P_- - P_+)]}{8C^2_W
S^2_W(s-m^2_Z)(s-m^2_{Z'})}\Bigg
\}\left\{-4s+\frac{2}{3}\frac{K^2_{H^-_2
W^+}}{m^2_W}\right\},\label{eq.24}
\end{eqnarray}
where $v_e=-1+4S^2_W$, $a_e =-1$,
$v'_e=\frac{v_e}{\sqrt{4C^2_W-1}}$,
$a'_e=\frac{a_e}{\sqrt{4C^2_W-1}}$,
\begin{eqnarray}
e_Z=\frac{v \omega}{2S^2_W\sqrt{\omega^2+C^2_\theta
v^2}}\left[S_\theta C_\theta(u_{12}+\sqrt{3}u_{22}) + C^2_\theta
u_{42}\right],
\end{eqnarray}

\begin{eqnarray}
e'_Z=\frac{v \omega}{2S^2_W\sqrt{\omega^2+C^2_\theta
v^2}}\left[S_\theta C_\theta(u_{13}+\sqrt{3}u_{23}) + C_{2\theta}
u_{43}\right],
\end{eqnarray} and

\begin{eqnarray}
K_{H^- W^+}=\left[(s-m^2_{H^-_2} - m^2_W)^2 - 4 m^2_{H^-_2}
m^2_W\right]^{1/2}.
\end{eqnarray}
Here we use data as in previous item  and
$m_{Z'}=800GeV$\cite{longs}.  For the sake of convenience in
calculations, we take the upper value of the mixing angle
$(s_{\theta})_{max}=
0.08$.\\
\hspace*{0.5cm} In Fig.5 we have plotted the total cross-section
$\sigma_{(P_+, P_-)}$ as a function of the polarized coefficients
with $P_-,P_+\in [-1,1]$. The total cross-section $\sigma_{(P_+,
P_-)}$ increases when $P_-$ comes to -1, $P_+$ comes to +1 and it
gets a maximum value by $\sigma_{(P_+, P_-)} =2.65\times
10^{-6}pbarn$ for $P_- = -1$ and $P_+ = 1$, while the $e^+$, $e^-$
beams are not polarized then the total cross-section as $\sigma_0$
= 0.8$\times 10^{-6}pbarn$. The DCS as function of $cos\theta$ at
the fixed energy, typically $\sqrt{s}$ = $1TeV$, is shown in
Fig.6. The curve 1 with $P_- = P_+ = 0$ and the curve 2 with $P_-
= -1$, $P_+ = 1$. From Fig.6 we can see  the behavior of DCS is
similar as in Fig. 2. The Higgs mass dependence of the total
cross-section for the fixed energies as in previous item  are
shown in the figures, Fig.7 for $P_-=-1$, $P_+=1$ and  Fig. 8 for
$P_-= P_+=0$. The mass range is $200GeV \leq m_H \leq 800GeV$. As
we can see from the figures, the total cross-section decreases
rapidly as $m_H$ increases. In Fig.9 we have plotted the Higgs
mass dependences of the total cross-section at the fixed higher
energies, $\sqrt{s}$ = $1TeV$, $\sqrt{s}$ = $1.5 TeV$ and
$\sqrt{s}$ = $2 TeV$.  From Fig.9 we can see that at high values
of $\sqrt{s}$, the total cross-section decreases slowly, while at
lower energy ( $\sqrt{s}$ = $1TeV$) the total cross-section
decreases rapidly as $m_H$ increases and there is no difference
between three  lines at $m_H \simeq 750$ $GeV$. Based on the
results,  it shows that the cross -section of pair $H^-_2W^+ $ is
much smaller than those of pair production $H^\pm_2$ in the same
condition .{\it  We deduce that the direct production of
$H^{-}_2W^+$ is in general not expected to lead to easily
observable signals
in $e^+e^-$ collisions}.\\
 \hspace*{0.5cm}In the final state, $H^-_2$ can
decay with modes as follows \bea H^-_2 & \rightarrow & l \nu_l,
\hs \tilde{U} d_a,\hs D_\al \tilde{u}_a,\crn
 &\searrow& Z W^-,\hs Z^{\prime}W^{-},\hs XW^-,\hs ZY^-.
 \label{modes}\eea
 Note that in the effective approximation  $H^-_2$ is a bilepton.
Assuming that masses of the exotic quarks $(U, D_\al)$ are larger
than $M_{H^\pm_2}$, we come to the fact that, the hadron modes are
absent in  decay of the  charged Higgs boson. Due to that the
Yukawa couplings of $H_2^\pm l^\mp \nu$ are very small, the main
decay modes of the $H_2^\pm$ are in the second line of
(\ref{modes}).
 Because of the exotic $X,Y,Z'$ gauge
bosons are heavy, the coupling of a singly-charged Higgs boson
($H^\pm_2$) with the weak gauge bosons, $H^\pm_2 W^\mp Z$, may be
main. For more details, the readers can see in our early work\cite{haihiggs1}\\

\section{Conclusions}
\label{conclus} In conclusion  we have presented the economical
3-3-1 model. In this model  we have focused attention to the
singly-charged Higgs boson  with
 mass is estimated in range of few hundred $GeV$.
 Pair production of $H^{\pm}_2$ in high energy $e^+ e^-$ colliders with
polarization of $e^+$, $ e^-$ beams was studied in detail.
Numerical evaluation shows that if the mass of them is not too
heavy and at high degree of polarization then the production cross
section may give observable values at moderately high energies.
With the  high integrated luminosity $ L = 500$ $fb^{-1}$, the
number of events can be several thousand as expected. We have
considered pair production of $H^{-}_2W^+$ at high energy $e^+
e^-$ colliders. From our results, it shows that cross - sections
for their production at high energies are very small, much below
production cross sections of $H^{\pm}_2$, so that the direct
production of $H^{-}_2W^+$ is in general not
expected to lead to easily observable signals in $e^+e^-$ annihilation.\\
 \hspace*{0.5cm} Finally, we emphasize that the
electron-positron linear collider in the energy range between
$500$ $GeV$ and $1000$ $GeV$ is very important to precisely
test the SM and to explore the physics beyond it.\\

\section*{Acknowledgments}
One of the authors ( D. V. Soa ) expresses sincere gratitude to
the European Organization for Nuclear Research ( CERN ) for the
financial support.  He is also grateful to Prof. J. Ellis for the
hospitality during his stay at CERN. This work
was supported in part by National Council for Natural Sciences of Vietnam.\\[0.3cm]

\appendix
\section{Mixing matrix of neutral gauge bosons}
For the sake of convenience in practical calculations, we used the
mixing matrix
\be \left(%
\begin{array}{c}
  W_3 \\
  W_8 \\
  B \\
  W_4 \\
\end{array}%
\right) =\mathrm{U}\left(%
\begin{array}{c}
  A \\
  Z^1 \\
  Z^2 \\
  W'_{4} \\
\end{array}%
\right),
 \ee where\begin{widetext}
 \be U= \left(%
\begin{array}{cccc}
  s_W & c_\va c_{\theta'}c_W  & s_\va c_{\theta'}c_W  & s_{\theta'}c_W \\
  \\
  -\fr{s_W}{\sqrt{3}} & \fr{c_\va(s^2_W-3c^2_Ws^2_{\theta'})
  -s_\va\sqrt{(1-4s^2_{\theta'}c^2_W)(4c^2_W-1)}}{\sqrt{3}c_Wc_{\theta'}} &
  \fr{s_\va(s^2_W-3c^2_Ws^2_{\theta'})+
  c_\va\sqrt{(1-4s^2_{\theta'}c^2_W)(4c^2_W-1)}}{\sqrt{3}c_Wc_{\theta'}} & \sqrt{3}s_{\theta'}c_W \\
  \\
  \fr{\sqrt{4c^2_W-1}}{\sqrt{3}} & -\fr{t_W(c_\va\sqrt{4c^2_W-1}
  +s_\va\sqrt{1-4s^2_{\theta'}c^2_W})}{\sqrt{3}c_{\theta'}} & -\fr{t_W(s_\va\sqrt{4c^2_W-1}
 -c_\va\sqrt{1-4s^2_{\theta'}c^2_W})}{\sqrt{3}c_{\theta'}}  & 0 \\
 \\
  0 & -t_{\theta'}(c_\va\sqrt{1-4s^2_{\theta'}c^2_W}
  -s_\va\sqrt{4c^2_W-1}) & -t_{\theta'}(s_\va\sqrt{1-4s^2_{\theta'}c^2_W}
  +c_\va\sqrt{4c^2_W-1}) & \sqrt{1-4s^2_{\theta'}c^2_W} \\
  \\
\end{array}%
\right)\nn.\ee \end{widetext} Here we have denoted  $s_{\theta'}=
\fr{t_{2\theta}}{c_W\sqrt{1+4t^2_{2\theta}}}$. In approximation,
the angle $\va$ has to be  very small~\cite{haihiggs}\be
t_{2\va}\simeq-\fr{\sqrt{3-4s^2_W}[v^2+(11-14s^2_W)u^2]}{2c^4_W\om^2}.\ee

\begin{figure}[htbp]
\begin{center}
\includegraphics[width=8.5cm,height=6cm]{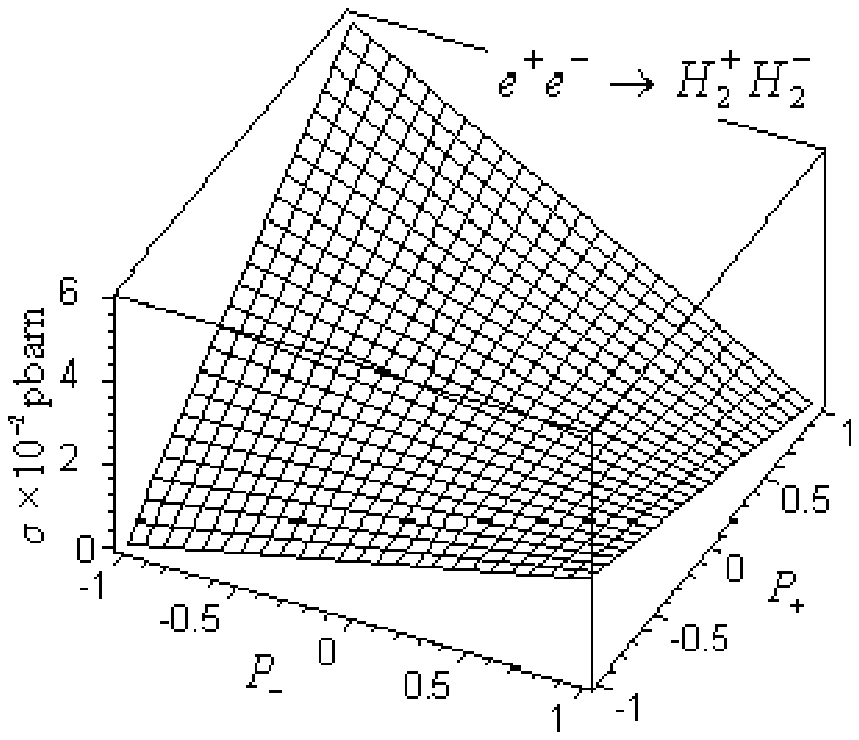}
\caption{\label{plseff}{The total cross-section of the process
$e^+ e^- \rightarrow H^{\pm}_2$ as function of polarized
coefficients.}} The collision energy is taken to be  $\sqrt{s} =
1$$TeV$, and Higgs  mass $m_{H^{\pm}_2} = 200$$GeV$.
\end{center}
\end{figure}
\begin{figure}[htbp]
\begin{center}
\includegraphics[width=8.5cm,height=5cm]{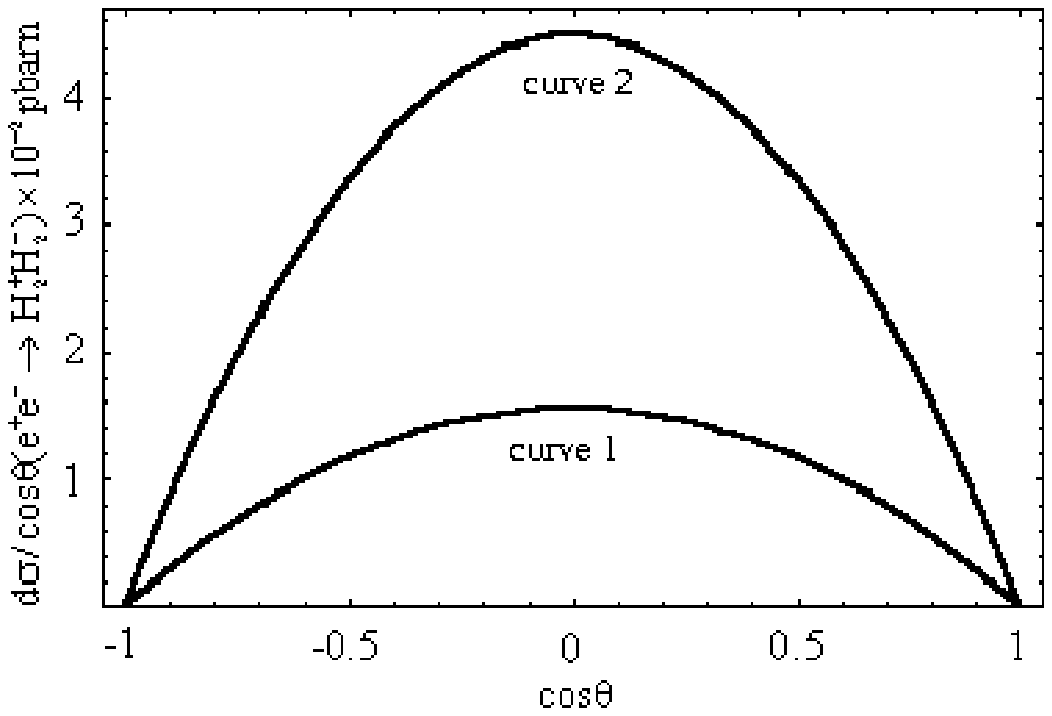}
\caption{\label{plseff}{The DCS of the process $e^+ e^-
\rightarrow H^{\pm}_2$ as function of cos$\theta$. The collision
energy is taken to be  $\sqrt{s} = 1$$TeV$,}} and Higgs  mass
$m_{H^{\pm}_2} = 200$$GeV$. The curve 1 for  $P_{-}= 0, P_{+}=0$,
and the curve 2 for $P_{-}=-1, P_{+}=1$.
\end{center}
\end{figure}

\begin{figure}[htbp]
\begin{center}
\includegraphics[width=8.5cm,height=5cm]{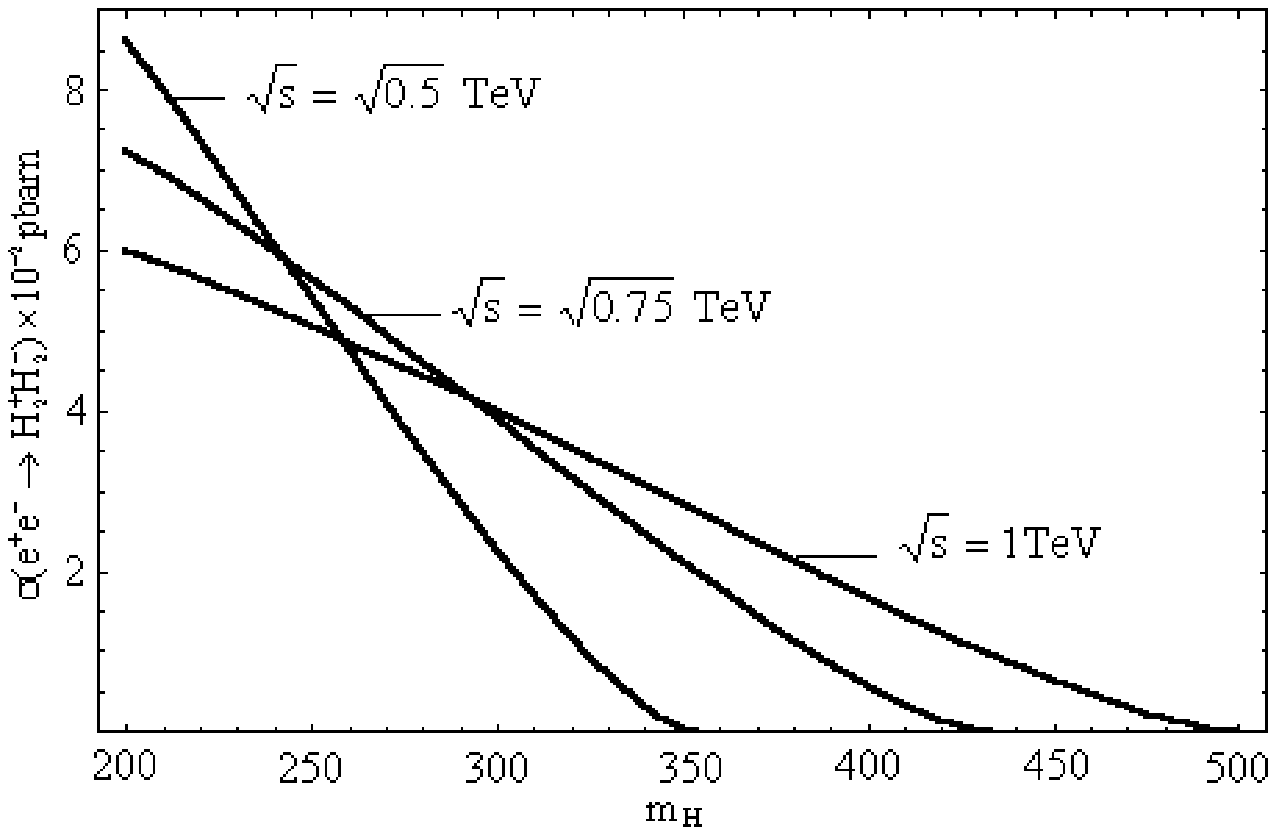}
\caption{\label{plseff}{The total cross-section of the process
$e^+ e^- \rightarrow H^{\pm}_2$ as function of $m_H$ for
$P_{-}=-1$ and $ P_{+}=1$.}}The collision energies are taken to be
$\sqrt{s} = 0.5 $$TeV$, $\sqrt{s} = 0.75$$TeV$, and $\sqrt{s} =
1$$TeV$, respectively.
\end{center}
\end{figure}

\begin{figure}[htbp]
\begin{center}
\includegraphics[width=8.5cm,height=5cm]{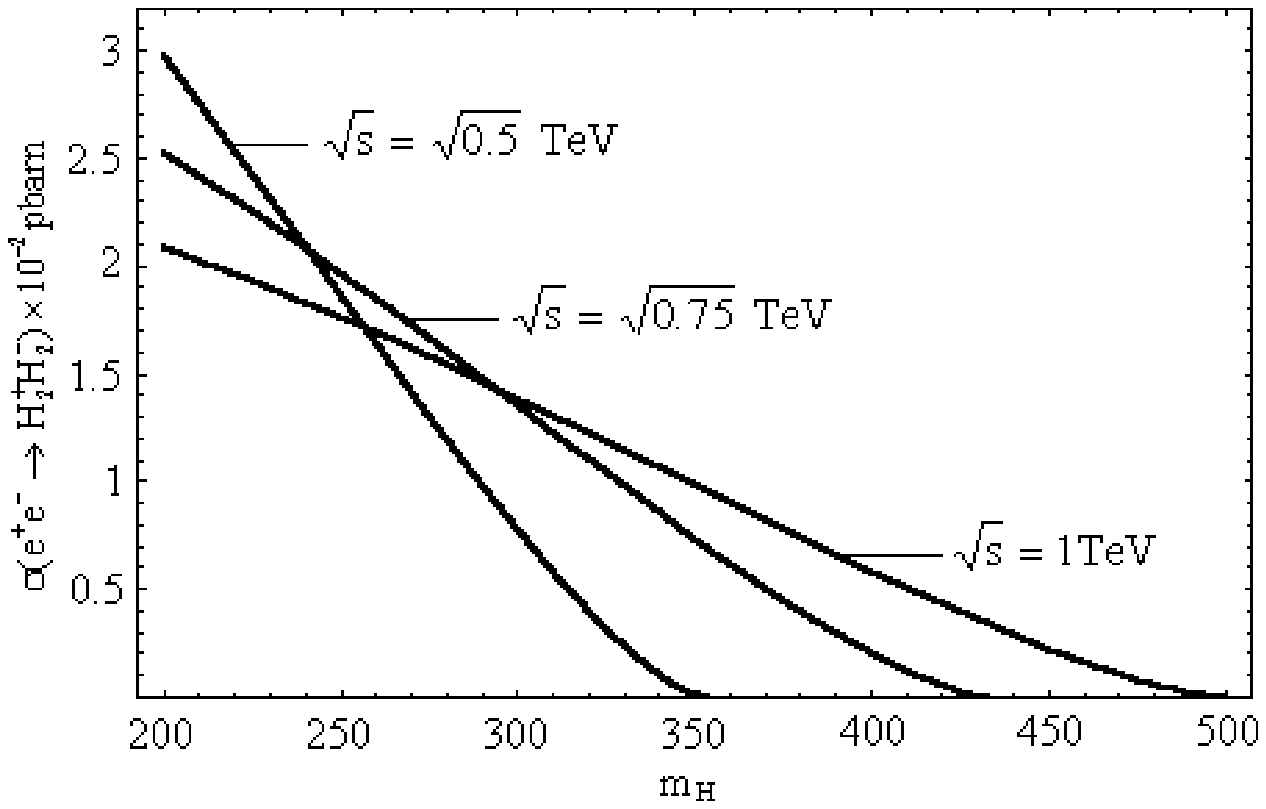}
\caption{\label{plseff}{The total cross-section of the process
$e^+ e^- \rightarrow H^{\pm}_2$ as function of $m_H$ for $P_{-}=
P_{+}=0$.}}The collision energies are taken to be $\sqrt{s} = 0.5
$$TeV$, $\sqrt{s} = 0.75$$TeV$, and $\sqrt{s} = 1$$TeV$,
respectively.
\end{center}
\end{figure}

\begin{figure}[htbp]
\begin{center}
\includegraphics[width=7.5cm,height=5cm]{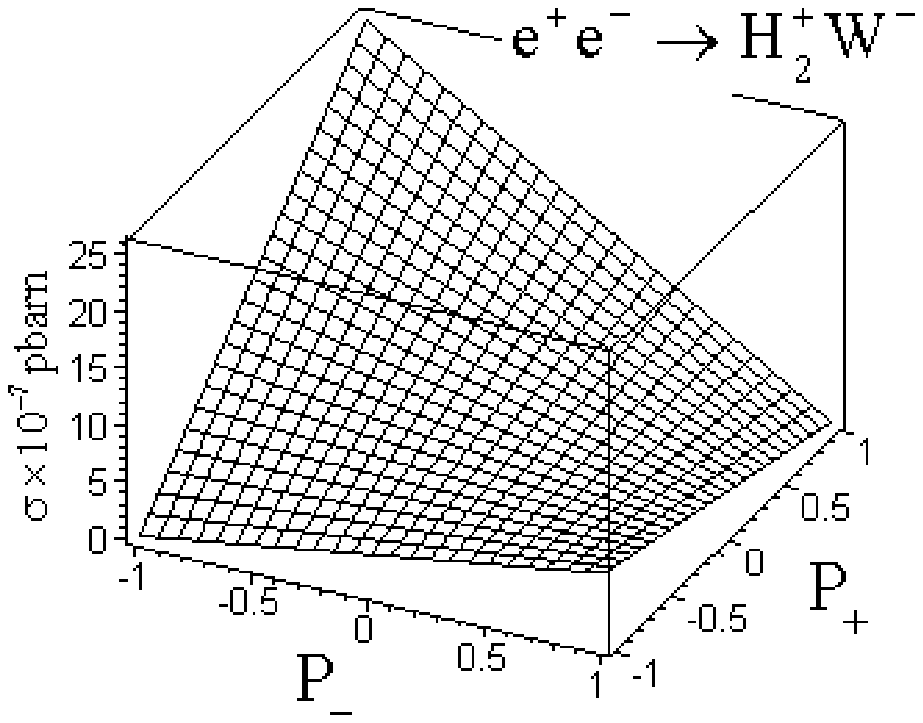}
\caption{\label{plseff}{The total cross-section of the process
$e^+ e^- \rightarrow H^{-}_2W^+ $ as function of polarization
coefficients.}} The collision energy is taken to be  $\sqrt{s} =
1$$TeV$, and Higgs mass $m_{H^{\pm}_2} = 200$$GeV$.
\end{center}
\end{figure}

\begin{figure}[htbp]
\begin{center}
\includegraphics[width=7.5cm,height=4.5cm]{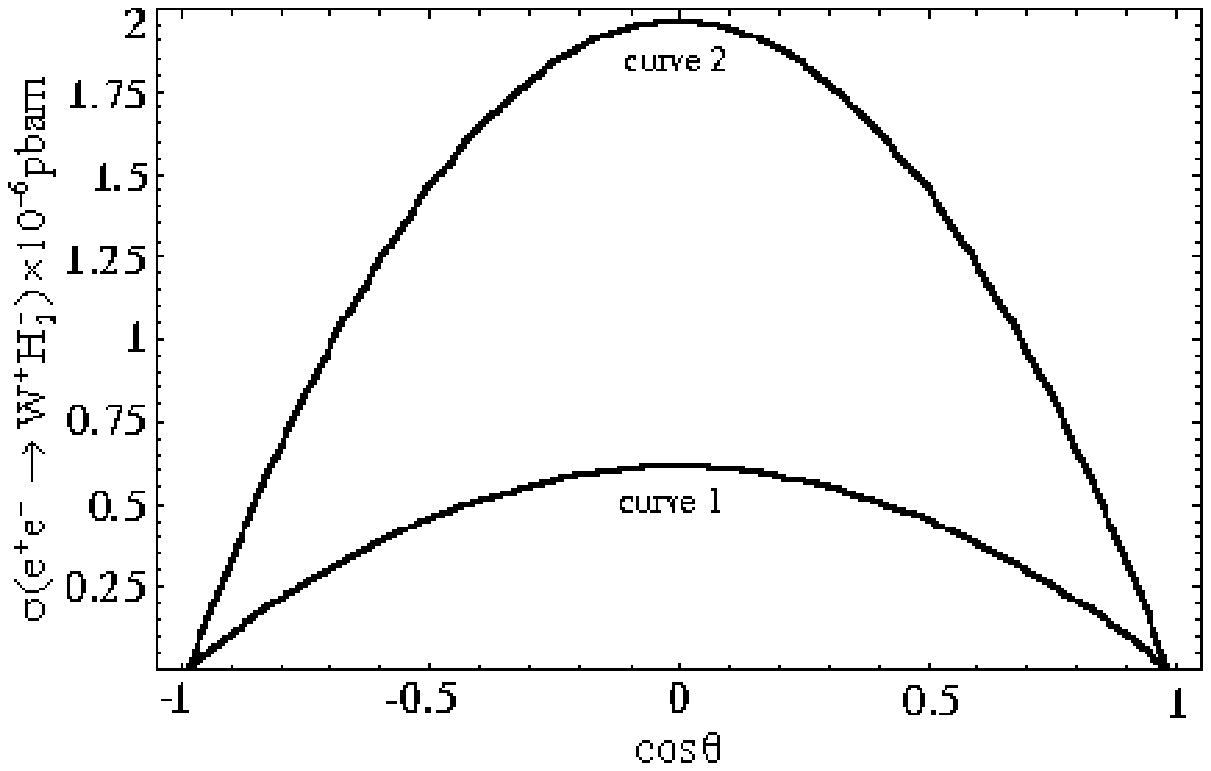}
\caption{\label{plseff}{The DCS as function of the process $e^+
e^- \rightarrow H^{-}_2W^+ $ of cos$\theta$. The collision energy
is taken to be  $\sqrt{s} = 1$$TeV$,}} and Higgs  mass
$m_{H^{\pm}_2} = 200$$GeV$. The curve 1 for $P_{-}= 0, P_{+}=0$,
and the curve 2 for $P_{-}=-1, P_{+}=1$.
\end{center}
\end{figure}

\begin{figure}[htbp]
\begin{center}
\includegraphics[width=7.5cm,height=4.5cm]{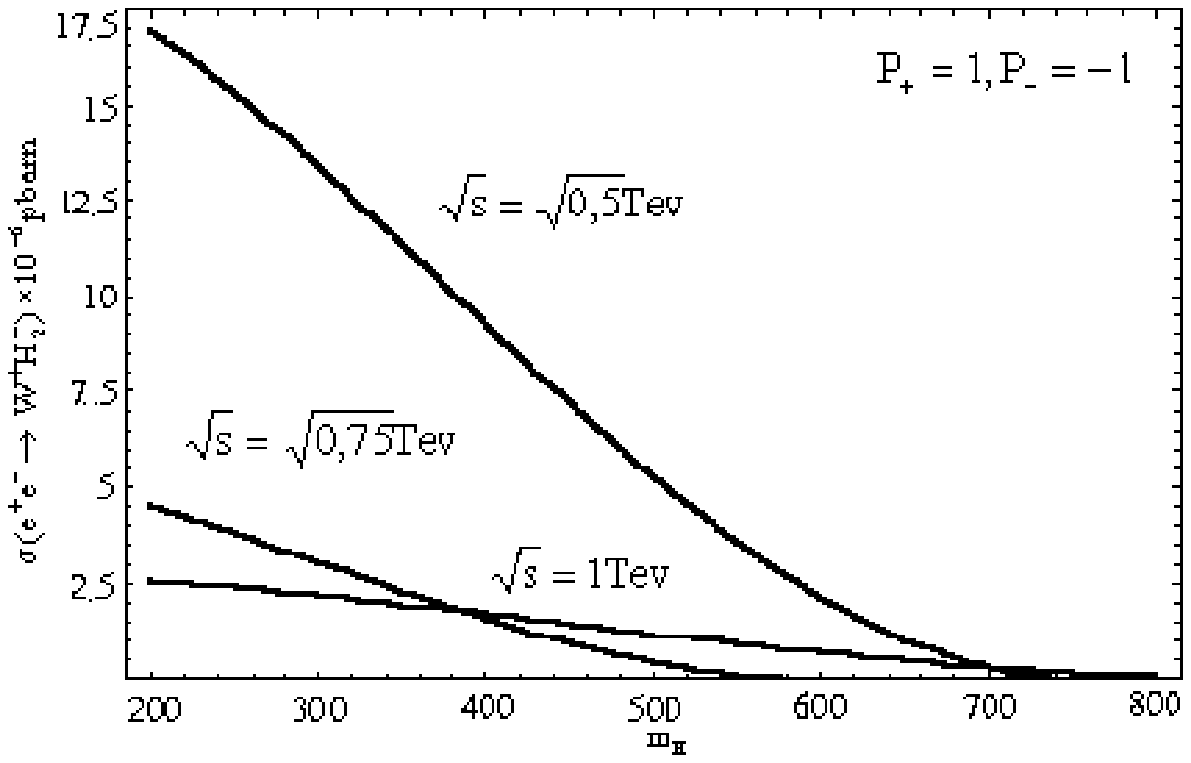}
\caption{\label{plseff}{The total cross-section of the process
$e^+ e^- \rightarrow H^{-}_2W^+ $  as function of $m_H$ for
$P_{-}=-1, P_{+}=1$.}} The collision energies are taken to be
$\sqrt{s} = 0.5 $$TeV$, $\sqrt{s} = 0.75$$TeV$, and $\sqrt{s} =
1$$TeV$, respectively.
\end{center}
\end{figure}

\begin{figure}[htbp]
\begin{center}
\includegraphics[width=7.5cm,height=4.5cm]{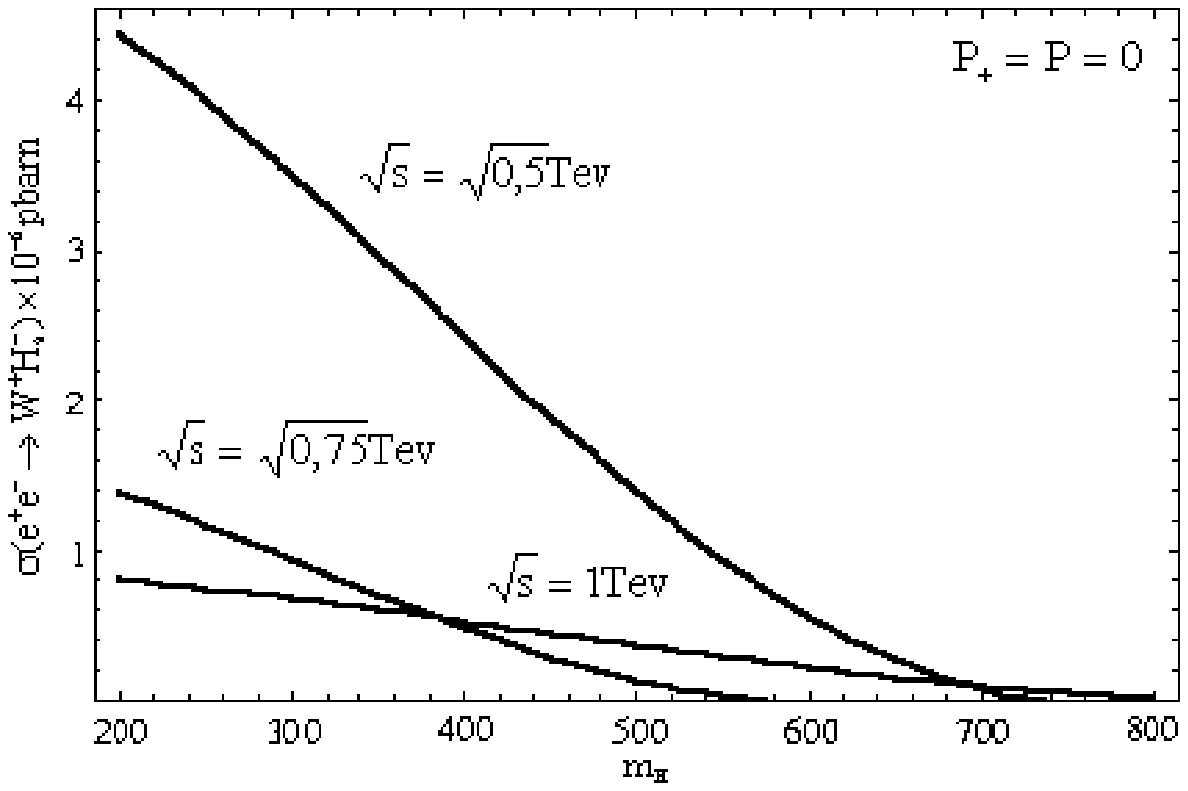}
\caption{\label{plseff}{The total cross-section of the process
$e^+ e^- \rightarrow H^{-}_2W^+ $ as function of $m_H$ for $P_{-}=
P_{+}=0$.}}  The collision energies are taken to be $\sqrt{s} =
0.5 $$TeV$, $\sqrt{s} = 0.75$$TeV$, and $\sqrt{s} = 1$$TeV$,
respectively.
\end{center}
\end{figure}

\begin{figure}[htbp]
\begin{center}
\includegraphics[width=7.5cm,height=4.5cm]{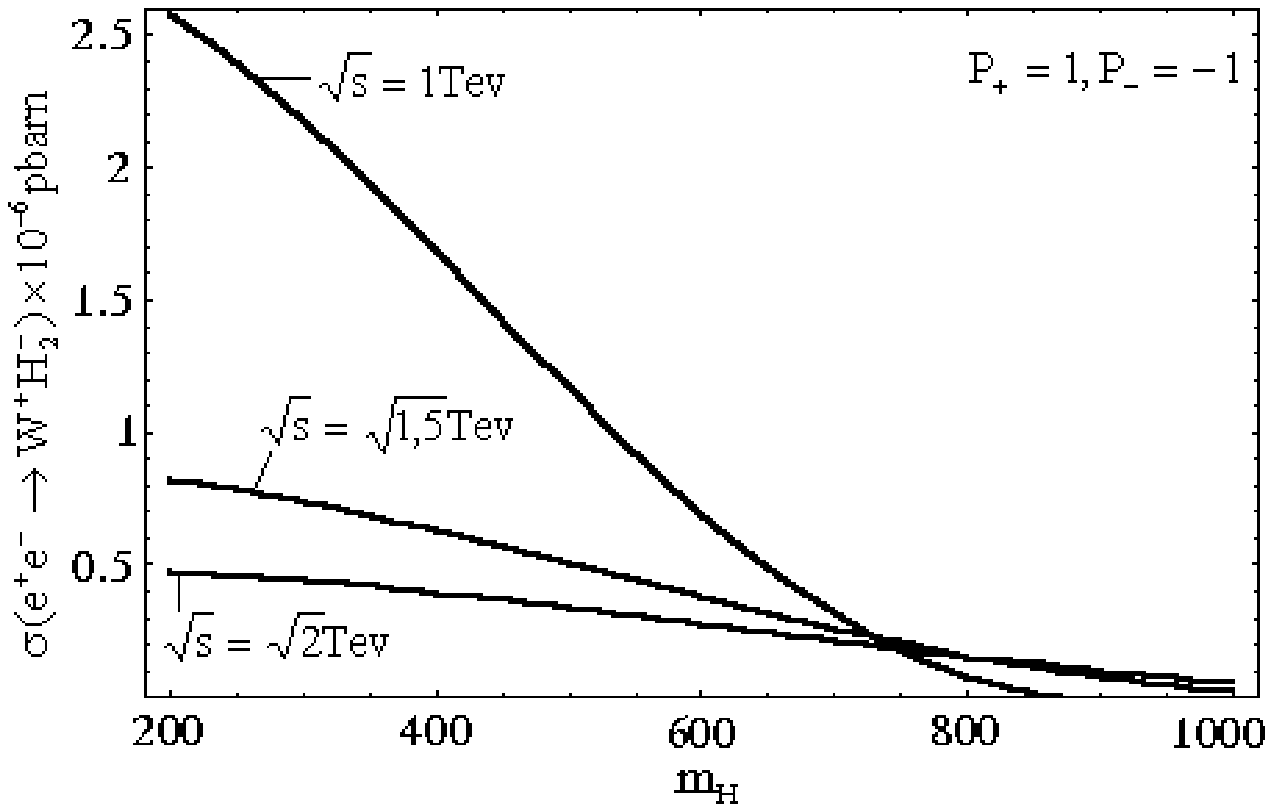}
\caption{\label{plseff}{The total cross-section of the process
$e^+ e^- \rightarrow H^{-}_2W^+ $ as function of $m_H$ for
$P_{-}=-1, P_{+}=1$.}} The higher collision energies are taken to
be $\sqrt{s} = 1$$TeV$, $\sqrt{s}= 1.5$$TeV$, and $\sqrt{s} =
2$$TeV$, respectively.
\end{center}
\end{figure}
\end{document}